# Emergence of long-range magnetic order from spin-glass state by tuning electron density in a stoichiometric Ga-based quasicrystal approximant

Farid Labib[1*], Shintaro Suzuki[1], Asuka Ishikawa[2], Takenori Fujii[3] and Ryuji Tamura[1]

[1] *Department of Material Science and Technology, Tokyo University of Science, Tokyo 125-8585, Japan*
[2] *Research Institute of Science and Technology, Tokyo University of Science, Tokyo 125-8585, Japan*
[3] *Cryogenic Research Center, The University of Tokyo, Bunkyo, Tokyo 113-0032, Japan*

This study reports the first observation of ferromagnetic (FM) order in the non-Au-based approximant crystals (ACs) using a novel approach whereby a total electron-per-atom ($e/a$) ratio of the spin-glass $Ga_{50}Pd_{36}Gd_{14}$ 1/1 AC is lowered by simultaneously substituting certain ratios of a tri-valent Ga and a zero-valent Pd by a mono-valent Au. The emergence of FM order by this method was confirmed via magnetic susceptibility, magnetization, and specific heat measurements. The findings of this study open up vast opportunities in developing long-range magnetic orders from stoichiometric ACs, quasicrystals, and even other RKKY compounds with spin-glass ground state.

## I. INTRODUCTION

Icosahedral quasicrystal ($i$QCs), as aperiodically ordered compounds, and their periodic counterparts, known as approximant crystals (ACs), are now commonplace in materials science since their first discovery in 1982 [1,2]. Tsai-type [3–5] $i$QCs and ACs, as a new class of strongly correlated electron system, showcase a wide range of physical properties such as a long-ranged antiferromagnetic (AFM) order that was first discovered in the binary $Cd_6RE$ ($RE$ = rare earth) 1/1 ACs [6]. However, a detailed study of their magnetism remained a major challenge due to the strong neutron absorbance of Cd which makes it unfavorable for neutron diffraction experiment. Such difficulty was later tackled by the discovery of various ternary Au-based ACs and recently $i$QCs with long-range magnetic orders [6–16]. As a significant accomplishment, a correlation between the electron per atom ($e/a$) ratio and the paramagnetic Curie-Weiss temperature ($\theta_w$) and magnetic ground states has been revealed in the Au-based ACs [10,17].

Since then, it was an open question whether the $e/a$ dependency of the $\theta_w$ and magnetic ground states is a unique characteristic of the Au-based systems or it is a universal behavior that can be expected in other alloy systems as well. The reason why this question has not been dealt with yet is that Tsai-type $i$QCs and ACs have rarely been reported in systems with major constituent elements other than Au and Cd. Besides, they are usually stable within a small compositional region in the phase diagram giving a very limited degree of freedom for their compositional tuning. Recently, a new non-Au-based Tsai-type 2/1 AC has been reported to form in the Ga-Pd-Tb system around a composition of $Ga_{50}Pd_{35.5}Tb_{14.5}$ [12]. This was followed by a comprehensive research done by the present authors to clarify phase stability, atomic structure, and magnetism of Ga-Pd-$RE$ ($RE$ = Gd, Tb, Dy, and Ho) 1/1 and 2/1 ACs [18–20].

In this article, we present a new experimental concept for obtaining long-range magnetic order in compounds with spin-glass (SG) ground state by introducing a new $e/a$ parameter space. We confirm that $e/a$ ratio is still a governing parameter in tuning magnetic ground states in non-Au-based ACs [17]. The findings of the present work open up vast future opportunities for realizing long-range magnetic order from "stoichiometric" compounds (not only QC-related ones but also other RKKY compounds) that exhibit SG-like ground state. These compounds, unlike Au-based ACs which are stabilized within a sizable single-phase region in the phase diagram, appear inside a narrow compositional area and thus are not tunable based on the conventional approach of simply changing the ratio of metallic species. The impact of such finding is even more highlighted when considering that a large number of reported Tsai-type compounds to date are stoichiometric with SG ground state, and thus are eligible for applying the proposed approach.

## II. EXPERIMENT

Polycrystalline Ga-Pd-Au-Gd 1/1 ACs were prepared by simultaneous substitution of Ga and Pd in the mother $Ga_{50}Pd_{36}Gd_{14}$ 1/1 AC by various contents of Au using an arc-melting technique. The Ga is kept as a major constituent element in the prepared samples. Nominal compositions used in the present study are listed in Table I. Powder X-Ray diffraction (XRD) was carried out for phase identification using Rigaku SmartLab SE X-ray Diffractometer with Cu-$K_\alpha$ radiation.

The dc magnetic susceptibility of the samples was measured under zero-field-cooled (ZFC) and field-cooled (FC) conditions using a superconducting quantum interference device (SQUID) magnetometer (Quantum Design, MPMS3) within $1.8 < T < 300$ K and in external dc fields up to $7 \times 10^4$ Oe. The ac magnetic susceptibility is also measured for some samples within $1.8 < T < 20$ K and under $H_{ac} = 1$ Oe and frequencies of 0.1, 1, 10, and 100 Hz.

## III. RESULTS

Figure 1 shows powder XRD patterns of the synthesized Ga-Pd-Au-Gd 1/1 ACs with various contents. The pattern at the bottom corresponds to the calculated one obtained from the refined atomic structure model of the Ga-Pd-Tb 1/1 AC [18], which is isostructural to the present compounds. The positions and intensity distributions of the experimental peaks are very well consistent with those of the calculated one indicating that the substituted Au has been dissolved into the structure of the mother Ga-Pd-Gd 1/1 AC resulting in single-phase quaternary 1/1 ACs with various Au contents spanning from 0 to 33 at.%. Figure 2 shows the variation of lattice parameters

…

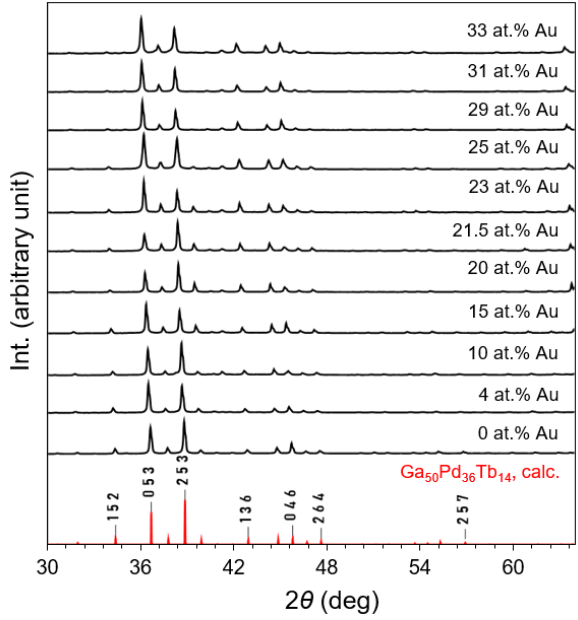

FIG. 1. Powder X-ray diffraction (XRD) patterns of the Ga-Pd-Au-Gd 1/1 ACs. The pattern at the bottom corresponds to the calculated one obtained from the refined atomic structure model of the Ga-Pd-Tb 1/1 AC [18].

estimated from Lebail analysis [21] of the powder XRD patterns versus Au content of the samples. Clearly, the lattice parameter expands linearly with the Au percentage, as expected from Vegard's law. This indicates the simultaneous replacement of Ga and Pd by Au within the structure. As a further confirmation, selected area electron diffraction (SAED) patterns are obtained from one of the synthesized samples containing 31 at.% Au (see the inset of FIG. 2 for the obtained pattern along [110]). The patterns along other main zone axes are provided in FIG. S1 of the Supplementary Material [22]. The patterns evidence the absence of any superlattice reflections which is consistent with the cubic lattice with the space group of $Im\overline{3}$. This further verifies the estimated lattice parameters from Lebail analysis where the space group of the samples was assumed to be $Im\overline{3}$.

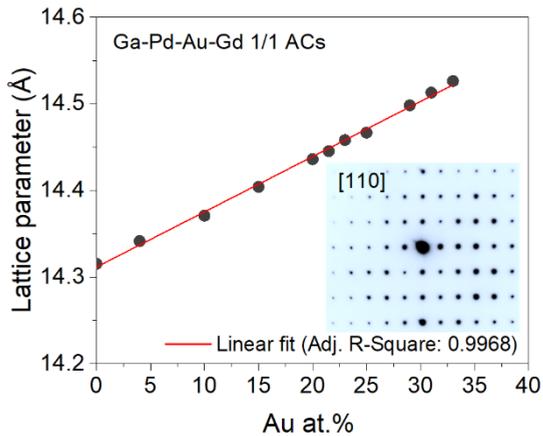

FIG. 2. Variation of lattice parameter of the synthesized Ga-Pd-Au-Gd 1/1 ACs obtained from Le Bail analysis as a function of their constituent Au content. The inset shows selected area electron diffraction (SAED) pattern obtained from $Ga_{34}Au_{31}Pd_{21}Gd_{14}$ along [110] axis.

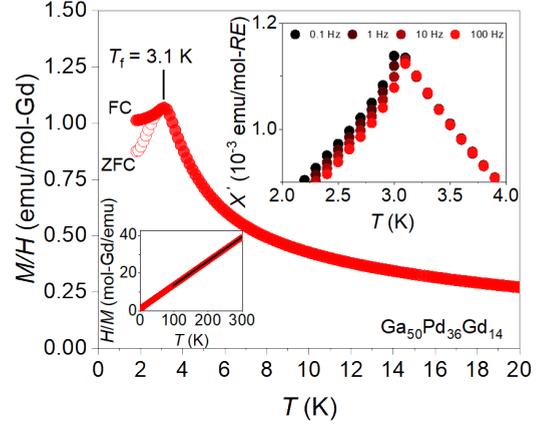

FIG. 3. Low-temperature magnetic susceptibility as a function of temperature under field-cooled (FC) and zero-field-cooled (ZFC) modes in the unsubstituted Ga-Pd-Gd 1/1 AC. The inset on the down-left part provides high-temperature inverse magnetic susceptibilities of the same compound. The inset on the up-right side shows a magnified view of the temperature-dependence of the in-phase ($\chi'_{ac}$) component in the ac magnetic susceptibility under frequencies spanning from 0.1 to 100 Hz.

Figure 3 provides temperature dependence dc magnetic susceptibility of the $Ga_{50}Pd_{36}Gd_{14}$ 1/1 AC within $1.8 < T < 20$ K in both field-cooled (FC) and zero-field-cooled (ZFC) modes, indicated by filled and unfilled circles, respectively. The figure evidences a bifurcation between ZFC and FC susceptibilities below $T = 3.1$ K, which is consistent with SG-like freezing behavior. The inset on the down-left part of the figure provides high-temperature inverse magnetic susceptibility ($H/M$) of the same compound within $1.8 < T < 300$ K showing a linear behavior well fitted to the Curie–Weiss law defined as:

$$\chi(T) = \frac{N_A \mu_{eff}^2 \mu_B^2}{3k_B(T-\theta_w)} + \chi_0 \qquad (1)$$

where $N_A$, $\mu_{eff}$, $\mu_B$, $k_B$, $\theta_w$, and $\chi_0$ represent the Avogadro number, effective magnetic moment, Bohr magneton, Boltzmann constant, Weiss temperature, and the temperature-independent magnetic susceptibility, respectively. The estimated $\theta_w$ obtained from a linear least-squares fitting of the inverse susceptibility data to the equation (1) in the range of 100 K $< T <$ 300 K and its extrapolation to the temperature axis is $-10.9\pm0.3$ K. Here, $\chi_0$ approximates zero. The obtained $\mu_{eff}$ is $8.0\pm0.2$ $\mu_B$ which is close to the calculated value of 7.94 for free $Gd^{3+}$ defined as $g\sqrt{J(J+1)}\mu_B$ ($g$ and $J$ stand for the Landé $g$-factor and total magnetic angular moment, respectively) [23] suggesting the localization of the magnetic moments on $Gd^{3+}$.

As a further confirmation of the SG state, ac magnetic susceptibility is measured. The inset on the up-right corner of FIG. 3 provides temperature-dependence of the resultant in-phase ($\chi'_{ac}$) component under selected frequencies ($f$) spanning from 0.1 to 100 Hz. The corresponding out-of-phase ($\chi''_{ac}$) component is shown in FIG. S2 of the Supplementary Material [22]. As seen, by rising the $f$, the position of the cusp in the $\chi'_{ac}$ component shifts ~ 1.5 % to higher temperatures while its magnitude lessens. The corresponding $\chi''_{ac}$ shows a small jump from a zero background below the freezing temperature ($T_f$) at $T = 3.1$ K. These results safely point out SG-like freezing behavior in the $Ga_{50}Pd_{36}Gd_{14}$ 1/1 AC. The mere 1.5 % rise of $T_f$ by three orders of magnitude change in $f$ originates from a weak response of the Heisenberg SGs to the measurement frequency change. It has been well demonstrated that spin dynamic (at low frequencies) is affected by the magnetic anisotropy and is less responsive to the measurement frequency change in Heisenberg SGs due to their exceptional isotropic behavior in the crystalline electric field (CEF) [24,25].

Figure 4 shows low-temperature (1.8 < $T$ < 20 K) magnetic susceptibility of the synthesized 1/1 ACs under $H_{dc}$ = 100 Oe. The high-temperature (1.8 < $T$ < 300 K) inverse magnetic susceptibility of the same compounds under $H_{dc}$ = 1000 Oe is provided in FIG. S3 of the Supplementary Material [22]. The ZFC and FC susceptibilities in FIG. 4 are represented by unfilled and filled circles, respectively. The estimated $\mu_{eff}$ and $\theta_w$ values are listed in Table I. The given uncertainties in the $\mu_{eff}$ and $\theta_w$ values in Table I correspond to standard deviations in the fitting over different temperature ranges within 100 < $T$ < 300 K as well as experimental resolution. All samples exhibit well-localized character for the $Gd^{3+}$ spins evidenced by the close agreement of their estimated Bohr magneton number with the theoretical value (see Table I). The magnitude of inverse magnetic susceptibility data in FIG. S3 of the Supplementary Material [22] shows a clear dependency on the Au content resulting in a continuous rise of $\theta_w$ from negative to positive values by increasing the Au content of ACs, which implies a change in the net magnetic interaction from AFM to FM.

What stands out in FIG. 4 is that the magnetic susceptibility of the ACs undergoes a sharp rise below a critical temperature $T_c$ in samples containing ≥ 20 at.% Au suggesting the occurrence of FM transition. Slight bifurcation of the ZFC and FC magnetizations below $T_c$ reflects small coercivity in the FM samples. In FIG. 5, the variation of $T_c$ (or $T_f$ in SG samples) estimated from the minimum of the first derivative of the FC magnetization, i.e., $d(M/H)/dT$ (see the inset of FIG. 5), is plotted as a function of Au at.% of the samples. Note that the Au contents shown in FIG. 5 are the nominal values. Unlike the lattice parameter that shows a monotonic trend in FIG. 2, the $T_c$ demonstrates a sudden jump above ~ 20 Au at.%, which is correlated with the onset of FM transition in the course of Au substitution.

Figure 6 displays field dependence magnetization (M–H) of the $Ga_{34}Au_{31}Pd_{21}Gd_{14}$ 1/1 AC at 1.8, 6, 10, 15, and 30 K up to 7 T. Ferromagnetic hysteresis loops are observed below $T_c$ ~ 10.5±0.1 K (as shown in Fig. S4 of the Supplementary Material [22]) even though the coercive field is very small implying a very soft magnetic behavior. The small coercivity in the present 1/1 ACs could partly be correlated to the weak CEF in the $Gd^{3+}$ due to its vanishing orbital component. At high magnetic fields, $M$ saturates, rather slowly, to nearly the full moment of a $Gd^{3+}$ free ion based on Hund's rule (i.e., 7.00 $\mu_B$/$Gd^{3+}$) suggesting the establishment of FM order. In the bottom-right inset of FIG. 6, the temperature dependence of the specific heat $C_p$ obtained from the same sample is provided in the temperature range of 0 – 20 K where the appearance of a pronounced anomaly is noticed confirming the FM order establishment. Based on the minimum of the first derivative d($C_p$)/d$T$ shown in the upper-left inset in Fig. 6, the transition takes place around 10.5±0.1 K. This marks the first example of FM state in the Tsai-type ACs reported to date whose main constituent element is not Au but Ga.

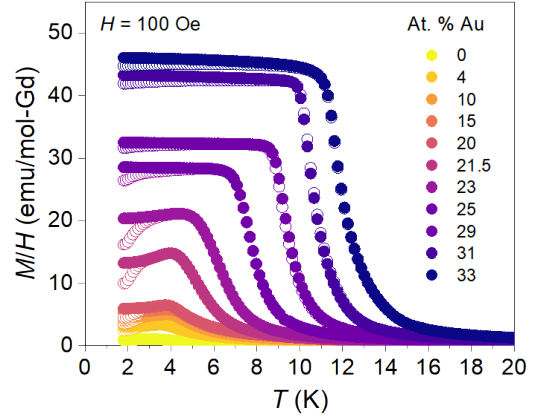

FIG. 4. Temperature-dependence of FC and ZFC magnetic susceptibilities (M/H) of the Ga-Pd-Au-Gd 1/1 ACs with various Au content within 1.8 < $T$ < 20 K.

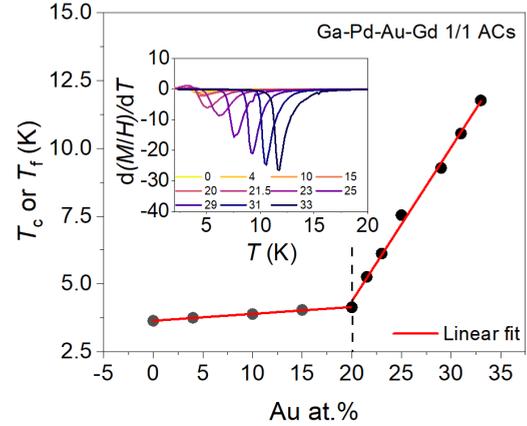

FIG. 5. Variation of critical temperature $T_c$ (or $T_f$ in SG samples) estimated from the minimum of the $d(M/H)/dT$ curves as a function of nominal Au content of the Ga-Pd-Au-Gd 1/1 ACs.

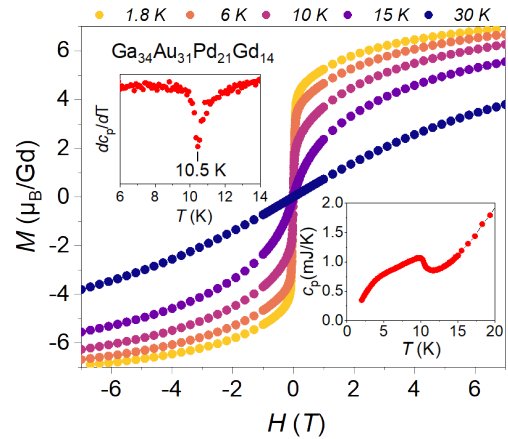

FIG. 6. Field-dependence of the magnetic susceptibility (M–H) of the $Ga_{34}Au_{31}Pd_{21}Gd_{14}$ 1/1 AC at 1.8, 6, 10, 15, and 30 K up to 7 T. The saturation occurs, though slowly, below $T_c$ = 10.5±0.1 K. The bottom-right inset shows the temperature dependence of the specific heat $C_p$ wherein an anomaly is observed confirming the occurrence of FM transition. The first derivative of $C_p$ in the upper-left inset suggests that the transition takes place at 10.5±0.1 K.

Table I. Nominal compositions, electron per atom ($e/a$), and magnetic properties of the Ga-Pd-Au-Gd 1/1 ACs

| Composition | $e/a$ | $\mu_{eff}$ ($\mu_B$) | $\theta_w$ (K) | $T_c$ or $T_f$ (K) |
|---|---|---|---|---|
| $Ga_{50}Pd_{36}Gd_{14}$ | 1.92 | 8.0±0.2 | −10.9±0.3 | 3.1 |
| $Ga_{48}Au_4Pd_{34}Gd_{14}$ | 1.90 | 8.1±0.2 | −6.2±0.3 | 3.7 |
| $Ga_{44}Au_{10}Pd_{32}Gd_{14}$ | 1.84 | 8.1±0.1 | −3.2±0.2 | 3.9 |
| $Ga_{42}Au_{15}Pd_{29}Gd_{14}$ | 1.83 | 8.0±0.2 | −2.7±0.3 | 4.0 |
| $Ga_{41}Au_{20}Pd_{25}Gd_{14}$ | 1.85 | 8.1±0.1 | −2.3±0.2 | 4.1 |
| $Ga_{40}Au_{21.5}Pd_{24.5}Gd_{14}$ | 1.83 | 8.2±0.2 | −0.7±0.1 | 6.1 |
| $Ga_{38.5}Au_{23}Pd_{24.5}Gd_{14}$ | 1.80 | 8.1±0.2 | +1.0±0.3 | 6.2 |
| $Ga_{37}Au_{25}Pd_{24}Gd_{14}$ | 1.78 | 8.0±0.1 | +3.3±0.1 | 7.5 |
| $Ga_{35}Au_{29}Pd_{22}Gd_{14}$ | 1.76 | 8.0±0.1 | +5.6±0.2 | 9.3 |
| $Ga_{34}Au_{31}Pd_{21}Gd_{14}$ | 1.75 | 8.1±0.2 | +7.1±0.4 | 10.5 |
| $Ga_{33}Au_{33}Pd_{20}Gd_{14}$ | 1.74 | 8.1±0.1 | +7.9±0.1 | 11.8 |

Since the number of valence electrons of Au is different from that of Ga, and Pd, the total electron density of the Au-substituted compounds should vary depending on their relative concentration (see Table I for $e/a$ values). Figure 7 plots electron per atom ($e/a$)-dependence of the normalized Weiss temperatures $\Theta_W$/$dG$ ($dG$ indicates de Gennes parameter expressed as $(g-1)^2 J(J+1)$ where $g_J$ and $J$ denote the Landé g-factor and the total angular momentum, respectively) of the present compounds and formerly studied Au-based ACs in various alloy systems [10,13–15,17]. The curves with different colors represent third-order polynomial fittings to the corresponding $\Theta_W$/$dG$ data. Here, Ga, Pd, and Au are assumed to be tri-, zero- and mono-valent [26], respectively. As shown, in the Ga-based ACs, the fitting curve crosses the baseline associated with $\Theta_W$/$dG$ = 0 at $e/a$ = 1.81, while in the Au-based systems, the baseline is intersected at about 1.88 < $e/a$ <1.91 (depending on the alloy system). A note of caution is due here considering the boundary between SG and FM ground states in the Au-based ACs that does not necessarily occur at $\Theta_W$/$dG$ = 0 but at slightly lower $e/a$ [17]. The observation of re-entrant SG behavior in the $Au_{67.2}Ge_{18.5}Gd_{14.3}$ 1/1 AC, the $e/a$ of which equals1.84 ($\Theta_W$/$dG$ = 0.77) [27], provides empirical evidence for the above claim. For a safe conclusion on the exact position of SG/FM boundary, further systematic investigation via, for example, specific heat measurement for the compounds with intermediate electron densities is required which falls beyond the scope of the present study. To express the uncertainty in the exact position of SG/FM boundary in FIG. 7, a region corresponding to 1.83< $e/a$ <1.87 is shaded in white. Relatively different $e/a$-dependency of the $\Theta_W$/$dG$ in Ga- and Au-based ACs in FIG. 7 may originate from different electron band structures in the two systems and/or non-zero valence electron number of Pd (in the present article, the valence electron number of Pd is assumed to be 0).

One way to explain the rise of $\Theta_W$ from negative to positive values by reducing $e/a$ is through considering the RKKY interaction which is proportional to $f(x)=(-x\cos x+\sin x)/x^4$; $x = 2k_F r$ and $dG$, with $k_F$ and $r$ corresponding to Fermi wave vector and the distance between two spins, respectively. Under the free-electron approximation, the $k_F$ is defined as $(\frac{3\pi^2 N}{V})^{1/3}$ where $N/V$ denotes the concentration of free electrons. Therefore, provided that $r$ remains unchanged during the Au substitution, dependency of $\Theta_W$/$dG$ on $e/a$ through $f(2k_F r)$ can be inferred. The range of the allowed e/a by the present compounds, however, is not wide enough to observe the decline of $\Theta_W$/$dG$ at the low electron density end in FIG. 7.

## CONCLUSION

In conclusion, we reported the emergence of FM order in the Tsai-type ACs whose main constituent element is not Au but Ga. We proposed a new experimental concept (i.e., introducing a new $e/a$ parameter space) in tuning the magnetic properties of Ga-Pd-Gd 1/1 AC from SG to FM by simultaneous substitution of tri-valent Ga and zero-valent Pd by a mono-valent Au while keeping the Gd percentage constant. The establishment of FM order was confirmed by magnetic susceptibility and specific heat measurements. The estimated Weiss temperature of the $Ga_{50}Pd_{36}Gd_{14}$ 1/1 AC with $e/a$ of 1.92 is noticed to increase from −10.9±0.3 K to +7.9±0.1 K in the $Ga_{33}Au_{33}Pd_{20}Gd_{14}$ 1/1 AC ($e/a$ = 1.74). The approach used in the present study opens up vast opportunities for realizing long-range magnetic order from "stoichiometric" compounds (not only QC-related ones but also other RKKY compounds) that exhibit SG-like ground state and are not tunable through the conventional approach of simply changing the ratio of metallic species. The impact of this finding is more noticeable when considering that a large number of reported Tsai-type compounds to date are stoichiometric with SG ground state, and thus are eligible for applying the proposed approach. The work to this end is under progress at the moment.

A number of possible future studies may follow the findings of the present work. One is elucidating the magnetic structure of the FM Ga-Au-Pd-Gd 1/l AC. Whether or not it would be iso-structural to the ferrimagnetic Au-Si-Tb 1/1 AC, where a non-coplanar magnetic order has been reported [28], is an interesting open question that needs to be addressed. Another follow-up research could focus on elucidating the boundary between SG and FM ground states in various alloy systems via specific heat measurement. Moreover, with the purpose of attaining magnetic orders, it will be important to explore the applicability of the approach used in the present work on other known 1/1 ACs and $i$QCs with spin-glass behavior. Ag-In- [29], Ga-Pt-, In-Pd- [30], and Cd-Mg-based [31] ACs and $i$QCs could be potential candidates in this regard.

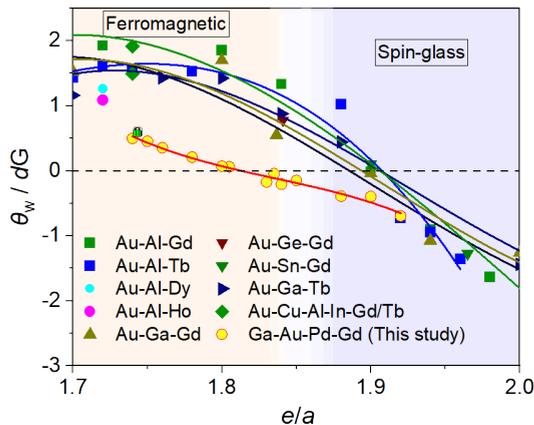

FIG. 7. Electron per atom ($e/a$)-dependence of the $\Theta_W$/$dG$ for the present Ga-Au-Pd-Gd 1/1 ACs as well as previously studied Au-based ACs in various alloy systems. The curves are polynomial fittings to the $\Theta_W$/$dG$ values. The fitting curves in the former and latter cross the baseline associated with $\Theta_W$/$dG$ = 0 at $e/a$ of 1.81 and 1.88 – 1.9, respectively. The white-shaded region in the background, corresponding to 1.83< $e/a$ <1.87, indicates the uncertainty in the exact position of SG/FM boundary.

## ACKNOWLEDGMENT

This work was supported by Japan Society for the Promotion of Science through Grants-in-Aid for Scientific Research (Grant Nos. JP19H05817, JP19H05818, JP19H05819 and JP21H01044).